# Semantic Web: Who is who in the field – A bibliometric analysis

**Ying Ding**[1]

*School of Library and Information Science, Indiana University Bloomington, IN, USA*

**Abstract**

**The Semantic Web is one of the main efforts aiming to enhance human and machine interaction by representing data in an understandable way for machines to mediate data and services. It is a fast-moving and multidisciplinary field. This study conducts a thorough bibliometric analysis of the field by collecting data from Web of Science (WOS) and Scopus for the period of 1960-2009. It utilizes a total of 44,157 papers with 651,673 citations from Scopus, and 22,951 papers with 571,911 citations from WOS. Based on these papers and citations, it evaluates the research performance of the Semantic Web (SW) by identifying the most productive players, major scholarly communication media, highly cited authors, influential papers and emerging stars.**

## 1. Introduction

The Web is experiencing tremendous changes in its function to connect information, people and knowledge, but also facing severe challenges to integrate data and facilitate knowledge discovery. The Semantic Web is one of the main efforts aiming to enhance human and machine interaction by representing data in an understandable way for machine to mediate data and services [1]. Recently, PriceWaterhouseCoopers [2] has predicted that Semantic Web technologies may revolutionize the entire enterprise of decision-making and information sharing. The profile of the Semantic Web has been further heightened by the Obama administration's new groundbreaking plan to initiate Semantic Web technologies to bring transparency to government activities [3]. Indeed, we see and hear the term "Semantic Web" almost everywhere.

Why is the Semantic Web becoming so popular? One obvious reason: the increasing needs of individuals and society to process information with efficiency, speed and comprehensiveness. This primary need addresses the vexing issue of the Web's over-flooded information. Ten years ago the Web largely contained documents,

---

[1] *Correspondence to: Ying Ding, School of Library & Information Science, Indiana University, 1320 East 10th Street, Herman B Wells Library, LI025, Bloomington, IN 47405, USA, dingying@indiana.edu*

# Ying Ding

allowing users to consume these documents and mine their nuggets in a timely and relatively straightforward fashion. But the explosively massive increase of websites has generated an information deluge, creating an often confusing and overwhelming format for gathering pertinent data within a reasonable period of time. For example, within the first few months of 2009 there has been an increase of 46 million websites [4]. The human capacity to read or consume this level of data is not possible to achieve in his lifetime. There is now a serious demand to distill documents into data that extracts core concepts from documents and represents them in a concise manner, such as RDF triples. A huge amount of documents can thus be shrunk to several data triples, which allows for easier consumption and retrieval. Yet even these improved cycles can go infinitively. As information deluge turns into data deluge, there is a need to add metadata to data, a process already in place. In turn metadata deluge will become another deluge, with no end in sight to this information abstraction. So while these abstracting processes can reduce the size of data and the burden for human use, they also create new challenges for data sharing and integration.

Another major issue for Web users is the problem of data sharing and integration [5]. If data is isolated somewhere as a silo, its usage and function can be significantly limited. As the world is becoming increasingly linked [6], the proper sharing of data has become essential to virtually all fields. Since data is represented in widely different syntax and semantics, the tasks of integrating data may be profoundly complex. One of the major missions for the database community is thus to find efficient ways to integrate data. But this remains as a remote goal where no "shortcuts" are available. Data stored in databases are structured data, while most data on the Web are unstructured[2]. Integrating and sharing data becomes more challenging, as what is called the current "bag-of-strings" nature of the Web does not facilitate connections that are machine-readable. These problems need to be addressed and solved. But there is no golden bullet. Semantic Web proposes technologies and methods that mainly address these two needs: how to add semantics to data and how to enable data integration [7].

Ontology, the backbone of the Semantic Web, is the formal representation of domain schemas. An ontology provides a shared vocabulary by modeling the semantics of data and representing them in markup languages proposed by the World Wide Web Consortium (W3C). W3C plays a major role in directing international efforts at specifying, developing and deploying standards for sharing information [8]. Semantically enriched data pave the crucial way to facilitate Web functionality and interoperability. Semantic Web technologies thus open up new possibilities for developing applications that work across the Web by modeling and linking data with best practices. They provide a fundamental infrastructure to create, represent and instantiate ontologies and metadata so as to enable intelligent retrieval and discovery. Semantic Web technologies continue to influence data sharing and management in various fields, such as Digital Library, Knowledge Management, Data Mining, Social Media, Electronic Commerce and Web Services [9].

Although Semantic Web is derived from the arena of Artificial Intelligence (AI), where ontological research can be traced back to early 1980s, the groundbreaking progress in this area started from the late 1990s or 2000, when significant funding was secured from the European Commission and United States to support these important innovations [5]. This paper uses citations and publications to illustrate this ten-year development in the Semantic Web field, with the special focus on semantics and ontology related research development. It is organized as the followings. Following this Introduction, Section 2 gives a brief history; Section 3 presents the research methods; Section 4 discusses the productivity and impact of this field, and Section 5 summarizes the results and addresses future research.

## 2. Brief history

Ten years ago nearly all the Semantic Web researchers could fit into one meeting room. They had to attend various conferences to explain the difference between "ontology" and "oncology," for the infrastructure and enabling methods/tools for Semantic Web were very unclear. Researchers still struggled with the migration of

---

[2] http://www.oracle.com/newsletters/information-insight/content-management/feb-07/forrester-future.html

**Ying Ding**

existing artificial intelligent methods to the Web, and tried to avoid yet-another "AI Winter" [10]. The brief history described here focuses on several "firsts" in the field: the first language, first conference, first journal and first foundation.

In Europe, we claimed that the first EU-funded Semantic Web project was OntoKnowledge (http://www.ontoknowledge.org/, 2000-2002). It was led by the Free University of Amsterdam. The major output of this project was the development of the OIL language which was not scheduled as a formal deliverable from the proposal. This project triggered the first meeting of EU and USA researchers in Aachen, Germany in August 2000. This meeting stressed the importance of the layered structure of OIL, and planned future EU funding for the community-formed Thematic Network for the Semantic Web (called the OntoWeb project, funded two years later by the EU). One month later, the second DAML and OIL meeting was held in Amsterdam. Three months later in December 2000, the DARPA Agent Markup Language Program officially announced that DAML+OIL was expected to be available that month, and in January 2001 its official version was released. DAML+OIL was later developed as OWL, which is currently the W3C standard and one of the key languages in the Semantic Web area.

The fundamental community-forming effort for the Semantic Web came from the OntoWeb project funded by EU from 2002-2004. The project created several "firsts" – the first conference, largely sponsored by the OntoWeb consortium, was held in Stanford in summer 2001. Named the Semantic Web Working Symposium. Afterwards, the conference was renamed as the International Semantic Web Conference and has been held annually thereafter in Europe, Asia and America in alternating years. Following the same pattern, the regional conferences were created. The first European Semantic Web Symposium was held in Greece 2004 and later on changed to the European Semantic Web Conference. The first Asia Semantic Web Conference was held in Beijing 2006. Of course, nowadays, Semantic Web related topics are mentioned in almost all the major computer science related conferences and broadly spread to conferences in other domains, such as, biology, chemistry, life science, medicine, library science and so on.

Creating an international journal for the field was planned in the deliverable of the OntoWeb project but was first discussed at the Dagstuhl Workshop on the Semantic Web in March 2000. The initial plan was to start the journal under the rubric of the Electronic Transactions on Artificial Intelligence (ETAI), which was published under the scientific patronage of the Royal Swedish Academy of Sciences and the European Coordinating Committee for Artificial Intelligence (ECCAI). This journal in the end found its home in Elsevier in 2003, named as *Journal of Web Semantics: Science, Services and Agents on the World Wide Web*. This journal grew with the community and received an impact factor of 3.023 from Journal Citation Report in 2009 published by Thomson Reuters. It is currently ranked as the 12th highest journal of 94 in the categories of Computer Science and Artificial Intelligence.

The first non-profit foundation was sponsored by the OntoWeb project and established in Amsterdam as "Stichting OntoWeb" (Stichting is the Dutch translation for "foundation") in 2001. The Foundation's objective is the advancement of research and development in the field of ontology and Semantic Web in general, and information exchange for knowledge management and electronic commerce in particular. Later on, this Stichting was moved to Karlsruhe and renamed the Semantic Web Science Association (SWSA). Now it supervises the organization of the International Semantic Web Conference series and other related conferences, workshops and summer schools and runs the *Journal of Web Semantics*.

At this ten-year juncture of the Semantic Web, it is now important to identify its current status, including who the major players are, such as, the most productive and highly cited authors, and the new driving forces. Since this area is moving fast and leading innovations on web engineering, data integration and service architecture, there is a pressing need to conduct research performance evaluation. This paper uses works published in this field to portray its research landscape.

## 3. Related work

Although various critical problems exist in bibliometric analysis as a method to evaluate research impact, such as database-related problems, inflated citation records, bias in citation rates and crediting of multi-author papers [11], it has been extensively applied over the past decades [12]. The basic approach is straightforwardly counting, such as how many times a particular paper has been cited [13]. Advanced techniques have been developed as well, such as author co-citation analysis [14], the h-index [15, 16], social network analysis [17, 18] and PageRank [19].

Recently, for example, Huang [20] collected publications associated with research on Obstructive Sleep Apnea (OSA) during the period of 1991-2006 from Web Of Science (WOS) to identify and predict the trends of publication output, journal patterns, country of publication, and authorship. Sorensen [16] applied citation analysis to post-1984 research on Alzheimer's Disease based on data from PubMed and WOS. Riikonen and Vihinen [13] examined the productivity and impact of more than 700 biomedical researchers in Finland from 1966 to 2000. Thijs and Glanzel [21] used different bibliometric indicators to profile European research institutes.

But there are not many available researches on using bibliometric methods to evaluate the field of Semantic Web, partially because it is still a young emerging field. Mika [22] and Mika, Elfring, and Groenewegen [23] conducted social network analysis for the Semantic Web research community based on researchers who have submitted publications or held an organizing role at the first, second and third International Semantic Web Conference (ISWC2002, ISWC2003 and ISWC2004) or the first Semantic Web Working Symposium in 2001. Their dataset contains 608 researchers. They compared the indegree, closeness, structural holes, publications and citations among these researchers and identified the core community and influential members. Zhao and Strotmann [24] used author co-citation analysis to detect school-of-thoughts for the XML field, which is quite broader than the Semantic Web field. As there is not a thorough citation analysis for Semantic Web research, this paper fills this gap by analyzing papers and citations produced in this field.

## 4. Method

For citation analysis, WOS and Scopus are the two major authorized databases [25]. But since 2007, WOS has excluded all the major computer science conference proceedings and put them to the ISI proceedings which are not part of WOS anymore[3]. Because Semantic Web is a young emerging multidisciplinary field, we place our focus especially on the semantics and ontology related research (as discussed in Introduction part), which form the core part of the Semantic Web field. In April 2009, "Semantic*" or "Ontolog*" have been used as the search terms to retrieve related publications and their citations from titles, keywords, and abstracts of papers in WOS and Scopus, with the restriction to the computer science related areas, including Library and Information Science[4]. The search query in WOS is TS[5]=(semantic* OR ontolog*) refined by subject areas related to computer science including theory and methods, artificial intelligence, information systems, software engineering, interdisciplinary, hardware and architecture, information science and library science, and cybernetics. There are 23,670 items identified. After excluding editorial materials, meeting abstracts and others, there are 22,951 articles remained. For Scopus, the search query is TITLE-ABS-KEY (semantic*) or title-abs-key(ontolog*) refined by subject areas in computer science, library and information science, and other related multidiscipline, which results in 46,029 items. After excluding corrections, conference review and other notes, there are 44,157 articles remained.

---

[3] http://isiwebofknowledge.com/media/pdf/cpci_faq.pdf

[4] Of course, there can be many other terms to retrieve related data in the Semantic Web field due to its multidisciplinary feature. But in this paper, we set our focus on research related to semantics and ontology (as addressed in the Introduction part), which are the crucial parts of the field. Other potential terms (e.g, RDF, XML, OWL, Linked Open Data, LOD, SPARQL, et al.) are therefore not included to retrieve data.

[5] TS in WOS include Title, Abstract, Author Keywords and Keywords Plus

**Ying Ding**

The main hypothesis for forming the search query for WOS and Scopus is that if this paper belongs to semantic web area, the authors should mention either "ontolog*" or "semantic*" in their title, keyword, or abstract. The reason why the semantic language terms are not included in the search query is that: 1) there are too many of them and they are still evolving, such as, XML, RDF, RDF-S, X-Query, SPARQL, RDFa, OWL, OIL, DAML+OIL, DAML, OWL-S, WSMO, WSML, GRIDDLE, SWRL, RIF, to name but a few. Also the OWL, OIL and DAML can lead to a large amount of noisy data, such as papers researching on OWL as an animal, or OIL as a product of oil industry. For example, Ian Horrock's most cited paper on OWL, in its title and abstract, there are semantic web and ontology mentioned. So if one paper never mentioned "ontolog*" or "semantic*" in title, abstract, or keyword, there is a high chance that this paper might not be directly related to the semantic web. So "ontolog*" or "semantic*" can be used to as search terms to capture the majority of papers published in the semantic web area[6].

In the end, there are 44,157 papers with 651,673 citations from Scopus covering 1975-2009, and 22,951 papers with 571,911 citations from WOS covering 1960-2009. We took these two datasets to analyze the research performance of the Semantic Web community. Semantic Web is a continuous development of the World Wide Web. The major progress of this field started from early 2000 when it gradually acquired major funding from European Commission and USA. In order to portray the details of this important phase, we divided the period of 2000-2009 into 2000-2004 and 2005-2009 to better outline its dynamic changes.

## 5. Results

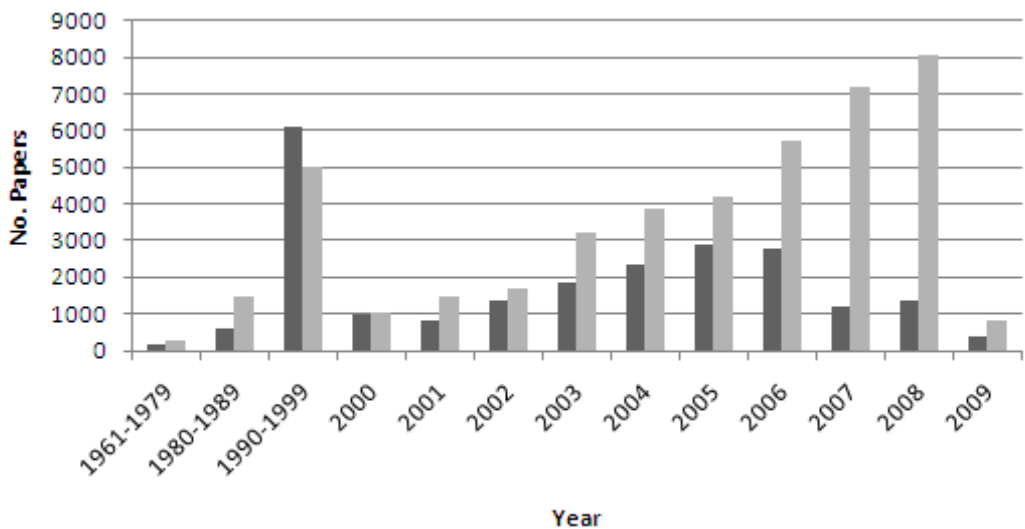

*Figure 1. No. of papers in WOS (dark grey) and Scopus (light Grey)*

---

[6] Just for testing purpose, in Feb 2010, there are around 7,600 articles in WOS having "OWL*" appeared either in title, keyword or abstract. Only less than 10% of them are related to semantic web. Among them, more than 95% have ontolog* or semantic* appeared either in title, keyword or abstract. Same testing for using „RDF*" as a search term for WOS, less than 30% of articles, which have RDF appears either in title, abstract or keyword, are related to semantic web area. Among them, more than 90% have ontolog* or semantic* appeared either in title, keyword or abstract.

# Ying Ding

There have been consistent increases in Semantic Web (SW) publications in Scopus (red column) to date, taking into consideration that 2009 data was downloaded in April 2009. In Scopus, the 2008 publications nearly doubled the amount of total paper published during 1990-1999. Since 2000, there has been an average yearly increase in publications of 31.7%. In WOS, however, these numbers significantly dropped in 2007 and 2008, due to the exclusion of conference proceedings from 2007 on, especially those coming from major Semantic Web events such as the International Semantic Web Conference, European Semantic Web Conference, Asian Semantic Web Conference and the World Wide Web Conference. Among the total number of SW papers in WOS and Scopus, 50% are conference papers.

## 5.1 Productivity

5.1.1  Journal/Conference

*Table 1. Major SW publication channels*

| WOS | | Scopus | |
|---|---|---|---|
| Journal/Conference | No. Paper | Journal/Conference | No. Paper |
| LECT NOTE COMPUT SCI | 7519 | Lecture Notes in Computer Science (including subseries Lecture Notes in Artificial Intelligence and Lecture Notes in Bioinformatics) | 9938 |
| LECT NOTE ARTIF INTELL | 2110 | Electronic Notes in Theoretical Computer Science | 1259 |
| THEOR COMPUT SCI | 741 | Lecture Notes in Artificial Intelligence (Subseries of Lecture Notes in Computer Science) | 860 |
| ACM SIGPLAN NOTICES | 355 | Theoretical Computer Science | 841 |
| BIOINFORMATICS | 290 | Proceedings of the National Conference on Artificial Intelligence | 342 |
| IEEE TRANS KNOWL DATA ENG | 278 | Data and Knowledge Engineering | 284 |
| DATA KNOWL ENG | 261 | Bioinformatics | 279 |
| INFORM COMPUT | 231 | IEEE Transactions on Knowledge and Data Engineering | 278 |
| ARTIF INTELL | 219 | Proceedings - International Conference on Data Engineering | 273 |
| EXPERT SYST APPL | 188 | Proceedings of the ACM Symposium on Applied Computing | 262 |
| J AMER MED INFORM ASSOC | 185 | Information and Computation | 252 |
| J LOGIC COMPUT | 175 | Artificial Intelligence | 252 |
| FUNDAM INFORM | 163 | Ruan Jian Xue Bao/Journal of Software | 247 |
| INFORM SYST | 154 | Fundamenta Informaticae | 224 |
| J LOGIC PROGRAM | 146 | Proceedings of the IEEE International Conference on Systems, Man and Cybernetics | 219 |
| SCI COMPUT PROGRAM | 144 | ICASSP, IEEE International Conference on Acoustics, Speech and Signal Processing - Proceedings | 216 |
| INFORM SOFTWARE TECHNOL | 143 | Information Systems | 215 |
| ACM TRANS PROGRAM LANG SYST | 142 | Expert Systems With Applications | 211 |
| ACTA INFORM | 136 | IEEE Transactions on Software Engineering | 210 |
| INFORM SCIENCES | 134 | Annual ACM Symposium on Principles of Programming Languages | 207 |

*Lecture Notes in Computer Science* and *Lecture Notes in Artificial Intelligence* are the two major publishing channels for SW papers, all of which are conference papers. This confirms that conference proceedings form the dominant publishing media reporting in the Semantic Web area. The top journals contributing to the publishing of SW papers are *Theoretical Computer Science*, *Bioinformatics*, *Data and Knowledge Engineering*, *IEEE transactions on Knowledge and Data Engineering*, *Information and Computation* and *Artificial Intelligence*. Most of the journals are in English, with one journal in Chinese, the *Ruan Jian Xue Bao/Journal of Software*.

5.1.2 Researchers

In WOS, the number of publications produced by the authors have been counted and ranked based on the first, second, and third author respectively. We present them in three time periods: 1960-2009, 2000-2004 and 2005-2009. For example, the first authors T. Eiter from Vienna Technical University, A. Brogi from University of Pisa, and H. Zhuge from Chinese Academy of Sciences are the top three most productive researches in 1960-2009. If we look at the recent period (2000-2009), H. Zhuge and T. Eiter keep their high productivity, while J. J. Jung from Yeungnam University, Korea emerges as a new star with 13 publications in 2005-2009, as does J. J. Alferes from University of Nova Lisboa, with 12 publications in 2000-2004 as first author.

*Table 2. Productive authors (WOS)*

| | 1960-2009 | | | 2000-2004 | | | 2005-2009 | | |
|---|---|---|---|---|---|---|---|---|---|
| R | First author | Second author | Third author | First author | Second author | Third author | First author | Second author | Third author |
| 1 | Eiter, T, 43 | Montanari, U, 32 | Montanari, U, 27 | Zhuge, H, 13 | Staab, S, 13 | Montanari, U, 9 | Eiter, T, 14 | Yang, Y, 9 | Vermeir, D, 9 |
| 2 | Brogi, A, 31 | Koutny, M, 20 | Prade, H, 16 | Alferes, JJ, 12 | Wu, ZH, 11 | Van der Hoek, W, 9 | Jung, JJ, 13 | Horrocks, I, 9 | Wu, ZH, 7 |
| | | | | Eiter, T, 12 | | | | | Castells, P, 7 |
| 3 | Zhuge, H, 26 | Wu, ZH, 19 | van der Hoek, W, 15 | | Vermeir, D, 11 | Terracina, G, 8 | Zhuge, H, 12 | Zhang, Y, 8 | |
| 4 | Bruni, R, 24 | Motta, E, 18 | Leone, N, 13 | Bruni, R, 11 | Montanari, U, 10 | Prade, H, 7 | Bruni, R, 10 | Chen, HJ, 8 | Li, Y, 6 |
| 5 | Antoniou, G, 22 | Vermeir, D, 18 | Geller, J, 13 | Bertino, E, 10 | de Boer, FS, 9 | Goble, C, 6 | Kim, W, 9 | Xu, BW, 8 | Di Sciascio, E, 6 |
| | | | Vermeir, D, 13 | | | | | Wu, ZH, 8 | Montanari, U, 6 |
| 6 | Jung, JJ, 21 | Lamma, E, 17 | | Fensel, D, 10 | Gelbukh, A, 9 | Elmagarmid, AK, 6 | Li, L, 9 | | Lu, JJ, 6 |
| 7 | Broy, M, 21 | | Subrahmanian, VS, 12 | Stojanovic, N, 9 | Chang, E, 8 | Hacid, MS, 6 | Horrocks, I, 9 | Li, X, 7 | Wang, XL, 6 |
| 8 | Bertino, E, 21 | Subrahmanian, VS, 17 | | Noy, NF, 9 | Motta, E, 8 | | Lee, J, 9 | Di Noia, T, 7 | Antoniou, G, 6 |
| | | Levi, G, 17 | Palamidessi, C, 12 | | Heckel, R, 8 | Pereira, LM, 6 | | | Wiklicky, H, 6 |
| 9 | Baldan, P, 20 | de Boer, FS, 16 | Terracina, G, 11 | Fan, JP, 8 | Serafini, L, 7 | van Harmelen, F, 6 | Baldan, P, 8 | Montanari, U, 7 | Sure, Y, 6 |
| 10 | Barbuti, R, 20 | | Sure, Y, 11 | Antoniou, G, 8 | Ling, TW, 7 | Ling, TW, 6 | Lukasiewicz, T, 8 | Motta, E, 7 | |
| 11 | Lee, J, 19 | Horrocks, I, 16 | Mello, P, 11 | Jung, JJ, 8 | Degano, P, 7 | Maedche, A, 6 | Zhang, Y, 8 | Tadeusiewicz, R, 7 | |
| | Greco, S, 19 | Staab, S, 16 | Lu, JJ, 11 | Maedche, A, 8 | Horrocks, I, 7 | Decker, S, 6 | Brogi, A, 8 | Parsia, B, 7 | |
| 12 | DEBAKKER, JW, 19 | Leone, N, 15 | Decker, S, 10 | Baldan, P, 8 | Pontelli, E, 7 | Ursino, D, 6 | Antoniou, G, 8 | Pontelli, E, 7 | Liu, L, 5 |
| 13 | | Gorrieri, R, 15 | Spyratos, N, 10 | Palopoli, L, 8 | Hahn, U, 7 | | Wang, Y, 8 | Jin, H, 7 | Leone, N, 5 |
| | | Pontelli, E, 15 | Meo, MC, 10 | Jacobs, B, 8 | Varadharajan, V, 7 | | | | Liu, J, 5 |
| 14 | Alferes, JJ, 18 | | | Mossakowski, T, 8 | Pereira, LM, 7 | Meseguer, J, 5 | Kim, J, 8 | | Steffen, M, 5 |
| 15 | Vogler, W, 18 | Gabbrielli, M, 14 | | | | Meersman, R, 5 | Lee, CS, 8 | Chang, E, 6 | Ho, CS, 5 |
| 16 | Zhang, Y, 18 | Zhang, Y, 14 | Perl, Y, 9 | | | Staab, S, 5 | | Reniers, MA, 6 | Zhang, L, 5 |
| 17 | Giacobazzi, R, 17 | Prade, H, 14 | Van Harmelen, F, 9 | Guarino, N, 7 | | Halper, M, 5 | Park, S, 7 | Jin, Z, 6 | Baik, DK, 5 |
| 18 | Dubois, D, 17 | Chang, E, 14 | Wiklicky, H, 9 | Bussler, C, 7 | Li, ML, 6 | Le, JJ, 5 | Lee, S, 7 | Smith, B, 6 | Chang, E, 5 |
| 19 | Borger, E, 16 | Xu, BW, 14 | Liu, J, 8 | Dau, F, 7 | Cimino, JJ, 6 | | Wang, P, 7 | Fokkink, W, 6 | |
| 20 | Corradini, A, 16 | Pereira, LM, 14 | Elmagarmid, AK, 8 | van Eijk, RM, 7 | Spyns, P, 6 | Tompits, H, 5 | Sanchez, D, 7 | Wand, Y, 6 | |
| | | | | | | Japaridze, | | | |
| | Aceto, L, 16 | | Meseguer, J, 8 | Ras, ZW, 7 | Perl, Y, 6 | Geller, J, 5 | G, 7 | Worring, M, 6 | |
| | Meseguer, J, 16 | | Steve, G, 8 | Benferhat, S, 7 | Niemela, I, 6 | Mongiello, M, 5 | Chen, Y, 7 | Hankin, C, 6 | |
| | Boreale, M, 16 | | Wu, ZH, 8 | Klein, M, 7 | Xu, BW, 6 | Sure, Y, 5 | Bry, F, 7 | Gasevic, D, 6 | |
| | | | Pereira, LM, 8 | Osorio, M, 7 | Kifer, M, 6 | Priami, C, 5 | Laird, J, 7 | Xu, D, 6 | |
| | | | Tarlecki, A, 8 | Boreale, M, 7 | Donini, FM, 6 | Meo, MC, 5 | Li, M, 7 | Li, HY, 6 | |
| | | | Liu, L, 8 | Bossi, A, 7 | Klein, M, 6 | Zavattaro, G, 5 | Liu, Y, 7 | Medeiros, CB, 6 | |
| | | | Rullo, P, 8 | Hunter, A, 7 | Meseguer, J, 6 | Snodgrass, RT | Kim, KY, 7 | | |
| | | | Di Sciascio, E, 8 | | Goble, C, 6 | Ma, FY | Heymans, S, 7 | | |
| | | | Ursino, D, 8 | | Palomar, M, 6 | Domingue, J | Bertini, M, 7 | | |
| | | | Mylopoulos, J, 8 | | Parsia, B, 6 | | Jovanovic, J, 7 | | |
| | | | Motta, E, 8 | | Zhang, WJ, 6 | | | | |
| | | | Goble, C, 8 | | | | | | |
| | | | Ling, TW, 8 | | | | | | |
| | | | Pugliese, R, 8 | | | | | | |

Note: number presents the number of publications. Some of the current Chinese names, such as Liu, L., Ding, L., can be the combination of different people, but it is beyond the scope of current research to differentiate author identities (same for other tables below).

Scopus contains all the excluded conference proceedings of WOS, and therefore has better coverage of the field. Within the total period, H. Zhuge, T. Eiter and J. J. Jung are the top three productive first authors. H. Zhuge maintains high productivity in 2000-2004 and 2005-2009, while J.J. Jung moves to the top one in 2005-2009 with 25 publications as first author, and T. Eiter keeps his third position in 2005-2009. E. Bertino from Purdue University and M. R. Naphade from University of Illinois are ranked as the second and third top productive first authors in 2000-2004.

*Table 3. Productive authors (Scopus)*

| | 1975-2009 | | | 2000-2004 | | | 2005-2009 | | |
|---|---|---|---|---|---|---|---|---|---|
| R | First author | Second author | Third author | First author | Second author | Third author | First author | Second author | Third author |
| 1 | Zhuge H., 44 | Yang Y., 30<br>Montanari U., 30 | Di Sciascio E., 47 | Zhuge H., 19 | Gelbukh A., 10 | Montanari U., 8<br>Smith J.R., 8 | Jung J.J., 25 | Di Noia T., 21 | Di Sciascio E., 21 |
| 2 | Eiter T., 37 | | Li X., 43 | Bertino E., 16 | Meseguer J., 9 | Terracina G., 8 | Zhuge H., 24 | Gasevic D., 19 | Esposito F., 18 |
| 3 | Jung J.J., 35 | Meseguer J., 28 | Motta E., 41 | Naphade M.R., 12 | Power J., 9<br>Dubois D., 9 | Staab S., 8 | Eiter T., 20 | Tadeusiewicz R., 18 | Motta E., 18 |
| 4 | Bertino E., 25<br>Yang Y., 25 | Staab S., 23<br>Motta E., 23<br>Di Noia T., 23 | Montanari U., 36 | Fensel D., 11 | Finin T., 9<br>Staab S., 9 | Elmagarmid A.K., 8 | Park S., 20<br>Yang Y., 20 | Hussain F.K., 18 | Baik D.-K., 16 |
| 5 | | | Esposito F., 33 | Noy N.F., 10 | | | | Jeong D., 17 | Hitzler P., 14 |
| 6 | Antoniou G., 24 | | Decker S., 30 | Benferhat S., 10 | | Hacid M.-S., 7 | Lee C.-S., 15 | Zhang D., 17 | Antoniou G., 13 |
| 7 | Park S., 23 | Tadeusiewicz R., 22<br>Parsia B., 22 | Domingue J., 27 | Eiter T., 10<br>Hunter A., 10 | Henderson-Sellers B., 8<br>Montanari U., 8 | Prade H., 7 | Dong H., 15<br>Lukasiewicz T., 15 | Dillon T.S., 17 | Tjoa A.M., 12<br>Domingue J., 12 |
| 8 | Broy M., | | Yang Y., | | Joshi A., | | | Montanari | |

# Ying Ding

| | | | | | | | | | |
|---|---|---|---|---|---|---|---|---|---|
| | 22 | | 26 | | Motta E., 8 | 6 | Fanizzi N., 14 | U., 16 | |
| 9 | Brogi A., 22 | Gomez-Perez A., 20 | Wiklicky H., 25 | Halpern J.Y., 9 | | Decker S., 6 | Ogiela L., 14 | Horrocks I., 15 | Sure Y., 11 |
| 10 | Baldan P., 21 | Del Bimbo A., 20 | Staab S., 25 | Jacobs B., 9 | Bertino E., 7 | Goble C., 6 | Bertini M., 14 | Meseguer J., 15 | Dillon T., 11 |
| 11 | Bergstra J.A., 21 | Dillon T.S., 20 | Baik D.-K., 25 | Ma Z.M., 9 | Goble C., 7 | Finin T., 6 | Baldan P., 13 | Parsia B., 15 | Wiklicky H., 11 |
| 12 | Halpern J.Y., 20 | Gasevic D., 19 | Sure Y., 24 | Wang Y., 9 | Serafini L., 7 | Van Harmelen F., 6 | Horrocks I., 13 | Motta E., 15 | Castells P., 11 |
| 13 | Barbuti R., 19 | Hussain F.K., 18 | Dillon T., 21 | Jung J.J., 8 | Joshi A., 7 | Studer R., 6 | Di Pierro A., 13 | D'Amato C., 14 | Vermeir D., 11 |
| 14 | Fensel D., 18 | Finin T., 18 | Dillon T.S., 19 | Li B., 8 | Parsia B., 7 | Ferrari E., 5 | Brogi A., 13 | Gomez-Perez A., 14 | Ogiela M.R., 11 |
| 15 | Benferhat S., 18 | Jeong D., 18 | Hitzler P., 19 | Kim W., 8 | De Boer F.S., 6 | Shah M., 5 | Antoniou G., 13 | Staab S., 14 | Dillon T.S., 10 |
| 16 | Di Pierro A., 18 | Horrocks I., 18 | Antoniou G., 19 | Stojanovic N., 8 | Vermeir D., 6 | Wu G., 5 | | Pan J.Z., 13 | Decker S., 10 |
| 17 | Horrocks I., 17 | Subrahmanian V.S., 18 | Palamidessi C., 18 | Power J., 8 | Varadharajan V., 6 | Ghafoor A., 5 | Sidhu A.S., 11 | Embley D.W., 12 | Wuwongse V., 10 |
| 18 | Bruni R., 17 | Embley D.W., 17 | Mylopoulos J., 16 | Chen Y., 8 | Musen M.A., 6 | Rastogi R., 5 | Ceravolo P., 11 | Del Bimbo A., 11 | Shi Y., 10 |
| 19 | Alpuente M., 17 | Hankin C., 16 | Tjoa A.M., 16 | Zhang D., 8 | Fournet C., 6 | Wiklicky H., 5 | D'Amato C., 10 | Lee S., 11 | Zhou Y., 10 |
| 20 | Corradini A., 17 | Peng Y., 16 | Vermeir D., 15 | Palopoli L., 8 | Venkatesh S., 6 | Pugliese R., 5 | De Bruijn J., 10 | Liu D., 11 | Ren F., 10 |
| | Bry F., 17 | Li S., 16 | Van Harmelen F., 15 | Lim J.-H., 8 | Gorrieri R., 6 | Harper R., 5 | Jovanovic J., 10 | Bielikova M., 11 | Liu Z., 10 |
| | Yager R.R., 17 | | Terracina G., 15 | | Mizoguchi R., 6 | Wang S., 5 | Bry F., 10 | Gugliotta A., 11 | Shi Z., 10 |
| | Lukasiewicz T., 17 | | Prade H., 15 | | Thiemann P., 6 | Snodgrass R.T., 5 | Mylonas P., 10 | Straccia U., 11 | |
| | Avron A., 17 | | | | Degano P., 6 | Venkatesh S. Le J., 5 | Huang W., 10 | | |
| | | | | | Parsons S., 6 | Subrahmanian V.S., 5 | Bruni R., 10 | | |
| | | | | | Zhou M., 6 | Van Der Hoek W., 5 | | | |
| | | | | | Mastroeni I., 6 | Ursino D., 5 | | | |
| | | | | | Ludascher B., 6 | Meseguer J., 5 | | | |
| | | | | | Sabry A., 6 | Sure Y., 5 | | | |
| | | | | | Huhns M.N., 6 | Maedche A., 5 | | | |
| | | | | | Bouguettaya A., 6 | Kim J., 5 | | | |
| | | | | | Grosky W.I., 6 | Ma W.Y., 5 | | | |
| | | | | | Chua T.-S., 6 | Heckel R., 5 | | | |
| | | | | | Pierce B.C., 6 | Brunie L., 5 | | | |
| | | | | | Heckel R., 6 | Domingue J., 5 | | | |
| | | | | | Tekalp A.M., 6 | Zhang L., 5 | | | |

Notes: some popular Asian names are deleted as many researchers can have the same names.

**5.2 Impact**

5.2.1 Highly cited Journals/conferences

In computer science, the major scholarly communication channel is shifting from journals to conferences[7]. Statistics on the Semantic Web, as one of the fast-moving subfields, show that the major highly cited channels are various conference proceedings published as *Lecture Notes in Computer Science* or *Lecture Notes in Artificial Intelligence*. In WOS (see Table 4), *Artificial Intelligence*, *Communication of the ACM* and *Theoretical Computer Science* journals are ranked the top three or four during these three periods. Looking at these top 20 highly cited journal/conferences, one finds that Semantic Web is closely related to artificial intelligence, computing theory, logic programming, database and bioinformatics.

*Table 4. Highly cited journals/conferences (WOS)*

| | 1960-2009 | | 2000-2004 | | 2005-2009 | |
|---|---|---|---|---|---|---|
| R | Journal/Conference | No. Cited | Journal/Conference | No. Cited | Journal/Conference | No. Cited |
| 1 | LECT NOTES COMPUT SC | 34015 | LECT NOTES COMPUT SC | 10604 | LECT NOTES COMPUT SC | 12706 |
| 2 | ARTIF INTELL | 6915 | ARTIF INTELL | 2164 | ARTIF INTELL | 2007 |
| 3 | THEOR COMPUT SCI | 5691 | LECT NOTES ARTIF INT | 1996 | LECT NOTES ARTIF INT | 1947 |
| 4 | COMMUN ACM | 5669 | THEOR COMPUT SCI | 1856 | COMMUN ACM | 1816 |
| 5 | LECT NOTES ARTIF INT | 4494 | COMMUN ACM | 1714 | THEOR COMPUT SCI | 1799 |
| 6 | J LOGIC PROGRAM | 3238 | INFORM COMPUT | 1206 | BIOINFORMATICS | 1181 |
| 7 | INFORM COMPUT | 3216 | J LOGIC PROGRAM | 1119 | INFORM COMPUT | 1048 |
| 8 | IEEE T SOFTWARE ENG | 3090 | IEEE T SOFTWARE ENG | 857 | IEEE T PATTERN ANAL | 1027 |
| 9 | ACM T DATABASE SYST | 2891 | IEEE T KNOWL DATA EN | 800 | NUCLEIC ACIDS RES | 1000 |
| 10 | J ASSOC COMPUT MACH | 2710 | ACM T PROGR LANG SYS | 778 | IEEE T KNOWL DATA EN | 987 |
| 11 | ACM T PROGR LANG SYS | 2567 | J ASSOC COMPUT MACH | 723 | IEEE T SOFTWARE ENG | 863 |
| 12 | IEEE T KNOWL DATA EN | 2289 | ACM T DATABASE SYST | 722 | ACM T PROGR LANG SYS | 705 |
| 13 | IEEE T PATTERN ANAL | 1842 | IEEE INTELL SYST APP | 653 | IEEE INTELL SYST APP | 662 |
| 14 | ACTA INFORM | 1617 | IEEE T PATTERN ANAL | 569 | DATA KNOWL ENG | 643 |
| 15 | ACM COMPUT SURV | 1478 | P ACM SIGMOD INT C M | 564 | J LOGIC PROGRAM | 595 |
| 16 | J ACM | 1431 | J LOGIC COMPUT | 462 | ACM T DATABASE SYST | 572 |
| 17 | P ACM SIGMOD INT C M | 1414 | J AM SOC INFORM SCI | 443 | J ASSOC COMPUT MACH | 542 |
| 18 | BIOINFORMATICS | 1400 | ACTA INFORM | 440 | J AM MED INFORM ASSN | 516 |
| 19 | J COMPUT SYST SCI | 1398 | DATA KNOWL ENG | 436 | VLDB J | 512 |
| 20 | NUCLEIC ACIDS RES | 1392 | ACM COMPUT SURV | 436 | INT J HUM-COMPUT ST | 511 |

Table 5 shows the top 20 highly cited journals or conferences from Scopus. There is no major difference between Table 4 and Table 5, where between them, *Lecture Notes in Computer Science*, *Communications of the ACM*, *Artificial Intelligence* and *Theoretical Computer Science* are ranked within the top three during these three periods. *Nature* and *Science* emerge within the top 20 in 2005-2009.

*Table 5. Highly cited journal/conference (Scopus)*

| | 1960-2009 | | 2000-2004 | | 2005-2009 | |
|---|---|---|---|---|---|---|
| R | Journal/Conference | No. Cited | Journal/Conference | No. Cited | Journal/Conference | No. Cited |
| 1 | Lecture Notes in Computer Science | 22923 | Lecture Notes in Computer Science | 6721 | Lecture Notes in Computer Science | 15176 |
| 2 | Communications of the ACM | 5913 | Theoretical Computer Science | 2511 | Communications of the ACM | 2983 |
| 3 | Theoretical Computer Science | 5564 | Communications of the ACM | 2228 | Artificial Intelligence | 2429 |
| 4 | Artificial Intelligence | 5069 | Artificial Intelligence | 2056 | IEEE Intelligent Systems | 2061 |

---

[7] http://isiwebofknowledge.com/media/pdf/cpci_faq.pdf

| 5 | IEEE Intelligent Systems | 2844 | Information and Computation | 1230 | Theoretical Computer Science | 2060 |
|---|---|---|---|---|---|---|
| 6 | Information and Computation | 2722 | IEEE Transactions on Software Engineering | 1018 | Bioinformatics | 1943 |
| 7 | Journal of the ACM | 2472 | Journal of the ACM | 995 | Lecture Notes in Artificial Intelligence | 1483 |
| 8 | Lecture Notes in Artificial Intelligence | 2371 | Journal of Logic Programming | 821 | Journal of the ACM | 1260 |
| 9 | Bioinformatics | 2203 | Lecture Notes in Artificial Intelligence | 800 | Computational Linguistics | 1188 |
| 10 | IEEE Computer | 1861 | IEEE Intelligent Systems | 771 | IEEE Transactions on Knowledge and Data Engineering | 1160 |
| 11 | IEEE Transactions on Software Engineering | 1842 | IEEE Computer | 677 | Information and Computation | 1142 |
| 12 | Computational Linguistics | 1679 | ACM Transactions on Programming Languages and Systems | 629 | Scientific American | 1113 |
| 13 | IEEE Transactions on Knowledge and Data Engineering | 1670 | ACM Computing Surveys | 574 | IEEE Computer | 1067 |
| 14 | ACM Computing Surveys | 1554 | Fuzzy Sets and Systems | 546 | IEEE Internet Computing | 988 |
| 15 | Scientific American | 1440 | IEEE Transactions on Knowledge and Data Engineering | 442 | IEEE Transactions on Software Engineering | 973 |
| 16 | Fuzzy Sets and Systems | 1389 | Electronic Notes in Theoretical Computer Science | 434 | Nucleic Acids Res | 908 |
| 17 | Data and Knowledge Engineering | 1353 | Science of Computer Programming | 432 | Data and Knowledge Engineering | 874 |
| 18 | IEEE Internet Computing | 1305 | Computational Linguistics | 432 | ACM Computing Surveys | 864 |
| 19 | ACM Transactions on Programming Languages and Systems | 1254 | ACM Transactions on Database Systems | 408 | Science | 858 |
| 20 | SIGMOD Record | 1212 | Acta Informatica | 390 | Nature | 833 |

5.2.2  Highly cited authors

The number of times authors or their works get cited can be used to measure the impact of their works on the community. Table 6 shows the top 20 highly cited authors based on 571,911 citations from WOS. In the whole period (1960-2009), R. Milner is ranked as the top one for his contribution of pi-calculus for mobile processes, M. Gelfond top two for his work of logic programming and non-monotonic reasoning, and C. A. R. Hoare top three for his Quicksort algorithm and Hoare Logic, which brought him the Turing Award in 1980. Sir Tim Berners-Lee, the inventor of the World Wide Web and Semantic Web, is ranked top four in the entire period, top two for 2000-2004 and top one for 2005-2009, which shows his increasing impact within the community. T. Gruber's ontology definition and his ontology engineering work are highly cited, causing him to be ranked as top three in 2000-2004. I. Horrocks's fundamental contribution to the Semantic Web languages, especially OWL, moves his rank up to top two in 2005-2009.

*Table 6. Highly cited authors in WOS*

| R | 1960-2009 | No. Cited | 2000-2004 | No. Cited | 2005-2009 | No. Cited |
|---|---|---|---|---|---|---|
| 1 | MILNER R | 2771 | MILNER R | 736 | BERNERSLEE T | 742 |
| 2 | GELFOND M | 1320 | BERNERSLEE T | 472 | HORROCKS I | 734 |
| 3 | HOARE CAR | 1308 | GRUBER TR | 445 | MILNER R | 655 |
| 4 | BERNERSLEE T | 1254 | FENSEL D | 418 | BAADER F | 529 |
| 5 | ABITEBOUL S | 1122 | DUBOIS D | 404 | GRUBER TR | 520 |

**Ying Ding**

| 6  | DUBOIS D    | 1088 | GELFOND M  | 400 | NOY NF         | 417 |
|----|-------------|------|------------|-----|----------------|-----|
| 7  | HORROCKS I  | 1057 | ABITEBOUL S| 384 | EITER T        | 403 |
| 8  | GRUBER TR   | 1051 | GUARINO N  | 359 | SALTON G       | 382 |
| 9  | ZADEH LA    | 1035 | COUSOT P   | 337 | GUARINO N      | 367 |
| 10 | GOGUEN JA   | 950  | EITER T    | 316 | ZADEH LA       | 348 |
| 11 | APT KR      | 911  | ABADI M    | 315 | GELFOND M      | 341 |
| 12 | BAADER F    | 897  | HORROCKS I | 313 | HOARE CAR      | 305 |
| 13 | PLOTKIN GD  | 895  | ZADEH LA   | 312 | FENSEL D       | 302 |
| 14 | HAREL D     | 895  | HAREL D    | 311 | DUBOIS D       | 300 |
| 15 | CARDELLI L  | 883  | SALTON G   | 299 | MAEDCHE A      | 292 |
| 16 | EITER T     | 879  | HOARE CAR  | 291 | ZHUGE H        | 267 |
| 17 | SALTON G    | 874  | CARDELLI L | 289 | ALUR R         | 264 |
| 18 | GUARINO N   | 839  | BAADER F   | 271 | ABADI M        | 264 |
| 19 | ABRAMSKY S  | 832  | ABRAMSKY S | 267 | VANDERAALST WMP| 264 |
| 20 | COUSOT P    | 828  | ALUR R     | 240 | MILLER GA      | 263 |

Citations in Scopus include all authors, making it possible to rank the cited authors based on first, second, and third author. In the total period (1960-2009), R. Milner, T. Berners-Lee and I. Horrocks are ranked as the top three highly cited first authors; J. Hendler, S. Staab, and H. Garcia-Molina are ranked as the top three highly cited second authors; O. Lassila, F. van Harmelen, and I. Horrocks are the top three highly cited third authors. In 2000-20004, R. Milner, T. Berners-Lee, and M. Abadi are the top three highly cited first authors; J. Hendler, V. Lifschitz, and H. Prade are top three highly cited second authors, and O. Lassila, F. van Harmelen and H. Prade are top three highly cited third authors. In 2005-2009, T. Berners-Lee, I. Horrocks and R. Milner are top three highly cited first authors; J. Hendler, S. Staab and I. Horrocks are top three highly cited second authors, and O. Lassila, F. van Harmelen and A. Joshi are the top three highly cited third authors.

*Table 7. Highly cited first, second and third authors (Scopus)*

| R | 1960-2009 | | | 2000-2004 | | | 2005-2009 | | |
|---|---|---|---|---|---|---|---|---|---|
|   | First author | Second author | Third author | First author | Second author | Third author | First author | Second author | Third author |
| 1 | Milner, R., 2182 | Hendler, J., 1937 | Lassila, O., 1538 | Milner, R., 916 | Hendler, J., 503 | Lassila, O., 364 | Berners-Lee, T., 1516 | Hendler, J., 1420 | Lassila, O., 1174 |
| 2 | Berners-Lee, T., 2033 | Staab, S., 944 | Van Harmelen, F., 614 | Berners-Lee, T., 511 | Lifschitz, V., 310 | Van Harmelen, F., 184 | Horrocks, I., 1126 | Staab, S., 705 | Van Harmelen, F., 429 |
| 3 | Horrocks, I., 1376 | Garcia-Molina, H., 807 | Horrocks, I., 437 | Abadi, M., 438 | Prade, H., 292 | Prade, H., 181 | Milner, R., 982 | Horrocks, I., 597 | Joshi, A., 305 |
| 4 | Salton, G., 1161 | Lifschitz, V., 781 | Walker, D., 433 | Dubois, D., 399 | Cousot, R., 267 | Walker, D., 164 | Gruber, T.R., 811 | Van Harmelen, F., 561 | Horrocks, I., 300 |
| 5 | Gruber, T.R., 1121 | Horrocks, I., 741 | Johnson, R., 393 | Abiteboul, S., 388 | Garcia-Molina, H., 262 | Johnson, R., 154 | Baader, F., 772 | Garcia-Molina, H., 507 | Hendler, J., 285 |
| 6 | Fensel, D., 1047 | Van Harmelen, F., 710 | Lenzerini, M., 371 | Fensel, D., 384 | Huang, T.S., 261 | Montanari, U., 143 | Salton, G., 724 | Parsia, B., 479 | Sheth, A., 269 |
| 7 | Guarino, N., 1022 | Huang, T.S., 659 | Joshi, A., 359 | Salton, G., 379 | Staab, S., 239 | Horrocks, I., 137 | Guarino, N., 634 | Patel-Schneider, P.F., 454 | Rahm, E., 256 |
| 8 | Baader, F., | Cousot, R., | Sheth, A., | Zadeh, | Meseguer, J., | Harper, | Noy, N.F., | Dumais, S.T., | Staab, S., |

# Ying Ding

|    | 1019 | 641 | 350 | L.A., 376 | 236 | R., 127 | 628 | 446 | 256 |
|----|------|-----|-----|-----------|-----|---------|-----|-----|-----|
| 9  | Zadeh, L.A., 1001 | Prade, H., 634 | Decker, S., 341 | Hoare, C.A.R., 371 | Montanari, U., 225 | Vianu, V., 126 | Fensel, D., 616 | Sattler, U., 434 | Lenzerini, M., 243 |
| 10 | Abadi, M., 923 | Dumais, S.T., 633 | Rahm, E., 340 | Abramsky, S., 337 | Pnueli, A., 224 | Lenzerini, M., 120 | Zhuge, H., 560 | Worring, M., 429 | Santini, S. |
| 11 | Hoare, C.A.R., 899 | Finin, T., 524 | Montanari, U., 338 | Cardelli, L., 336 | Parrow, J., 193 | Decker, S., 116 | Zadeh, L.A., 538 | Calvanese, D., 414 | Domingos, P., 232 |
| 12 | Dubois, D. | Bernstein, P.A., 522 | Prade, H., 334 | Cousot, P., 332 | Cardelli, L., 186 | Suciu, D., 109 | Maedche, A., 526 | Finin, T., 412 | Payne, T.R., 225 |
| 13 | Abramsky, S., 828 | Worring, M., 517 | Hendler, J., 333 | Guarino, N., 330 | Gorrieri, R., 176 | Rice, J., 107 | Paolucci, M., 488 | Lifschitz, V., 395 | Decker, S., 225 |
| 14 | Cousot, P., 822 | Montanari, U., 513 | Staab, S., 326 | Harel, D., 319 | Lenzerini, M., 168 | Wu, J., 102 | Eiter, T., 483 | Huang, T.S., 394 | Johnson, R., 223 |
| 15 | Abiteboul, S., 813 | Sattler, U., 512 | Santini, S., 311 | Alur, R., 309 | Dumais, S.T., 166 | Ullman, J.D., 100 | Hoare, C.A.R., 449 | Bernstein, P.A., 372 | Walker, D., 219 |
| 16 | Noy, N.F., 798 | Parsia, B., 507 | Domingos, P., 307 | Gelfond, M., 301 | Fikes, R., 162 | Widom, J., 99 | Alur, R., 441 | Musen, M.A., 336 | Finin, T., 211 |
| 17 | Gelfond, M., 795 | Meseguer, J., 507 | Fensel, D., 294 | Gruber, T.R., 256 | Fensel, D., 160 | Jacobson, I., 98 | Calvanese, D., 406 | Kawamura, T., 320 | Fensel, D., 206 |
| 18 | Alur, R., 784 | Lenzerini, M., 506 | Vianu, V., 290 | Meseguer, J., 249 | Grumberg, O., 159 | Steele, G., 97 | Abadi, M., 406 | Paolucci, M., 301 | Volz, R., 202 |
| 19 | Harel, D., 773 | Patel-Schneider, P.F., 505 | Harper, R., 288 | Horrocks, I., 240 | Helm, R., 156 | Booch, G., 90 | Gelfond, M., 405 | Cousot, R., 292 | Sattler, U., 195 |
| 20 | Eiter, T., 759 | Pnueli, A., 485 | Finin, T., 280 | Rui, Y., 237 | Bernstein, P.A., 146  Walker, D., 146 | Eker, S., 88 | Foster, I., 388  Gousot, P., 388 | Lenzerini, M., 280 | Boley, H., 194 |

### 5.2.3 Highly cited papers

Table 8 shows the ranks of highly cited papers in three different periods from WOS. T. Gruber's ontology paper has been consistently highly cited and ranked as the top one for all periods. M. Gelfond's stable model semantics for logic programming is ranked as the top two highly cited paper in 1960-2009 and top three in 2000-2004. A. van Gelder's well-founded semantics for general logic programs is ranked as the top three in 1969-2009 and the top two in 2000-2004. T. Berners-Lee, J. Hendler and O. Lassila's famous article about the vision of Semantic Web published in *Scientific American* is ranked as the top two highly cited paper in 2005-2009. M. Ashburner's Gene Ontology article is ranked as the top three highly cited paper in 2005-2009. Through examining the highly cited papers in this field, one sees a clear shift from its beginning as being heavy artificial intelligence-dominated with a focus on knowledge representation, logic programming and theory proving, to more data-driven practical approaches designed to realize the Semantic Web vision by converting the current document Web into a data Web. During 2005-2009, more papers from data mining, natural language processing and database are highly cited. Ontology forms the heart of the Semantic Web vision and approaches, and the community has accepted ontology definitions coming from T. Gruber. Ontology engineering is also moving from creating a theoretical foundation for ontology to the mapping of different ontologies. Ontology languages have slowly evolved from various logic languages derived from the core AI. Semantic Web services emerged in 2005-2009, mainly represented by OWL-S initiative (e.g., that of J. Hendler and S. Mcilarith).

*Table 8. Highly cited papers (WOS)*

|   | 1960-2009 | | 2000-2004 | | 2005-2009 | |
|---|-----------|---|-----------|---|-----------|---|
| R | paper | no. cited | paper | no. cited | paper | no. cited |
| 1 | GRUBER TR (1993), A TRANSLATION APPROACH TO PORTABLE ONTOLOGY SPECIFICATIONS, KNOWL ACQUIS, V5, P199 | 513 | GRUBER TR (1993), A TRANSLATION APPROACH TO PORTABLE ONTOLOGY SPECIFICATIONS, KNOWL ACQUIS, V5, P199 | 175 | GRUBER TR (1993), A TRANSLATION APPROACH TO PORTABLE ONTOLOGY SPECIFICATIONS, KNOWL ACQUIS, V5, P199 | 313 |

| 2 | GELFOND M (1988), THE STABLE MODEL SEMANTICS FOR LOGIC PROGRAMMING, P 5 INT C LOG PROGR, P1070 | 393 | VANGELDER A (1991), THE WELL-FOUNDED SEMANTICS FOR GENERAL LOGIC PROGRAMS, J ASSOC COMPUT MACH, V38, P620 | 108 | BERNERSLEE T (2001), THE SEMANTIC WEB - A NEW FORM OF WEB CONTENT THAT IS MEANINGFUL TO COMPUTERS WILL UNLEASH A REVOLUTION OF NEW POSSIBILITIES, SCI AM, V284, P34 | 238 |
|---|---|---|---|---|---|---|
| 3 | VANGELDER A (1991), THE WELL-FOUNDED SEMANTICS FOR GENERAL LOGIC PROGRAMS, J ASSOC COMPUT MACH, V38, P620 | 311 | GELFOND M (1988), THE STABLE MODEL SEMANTICS FOR LOGIC PROGRAMMING, P 5 INT C LOG PROGR, P1070 | 106 | ASHBURNER M (2000), GENE ONTOLOGY: TOOL FOR THE UNIFICATION OF BIOLOGY, NAT GENET, V25, P25 | 211 |
| 4 | DEERWESTER S (1990), INDEXING BY LATENT SEMANTIC ANALYSIS, J AM SOC INFORM SCI, V41, P391 | 265 | BERNERSLEE T (2001), THE SEMANTIC WEB - A NEW FORM OF WEB CONTENT THAT IS MEANINGFUL TO COMPUTERS WILL UNLEASH A REVOLUTION OF NEW POSSIBILITIES, SCI AM, V284, P34 | 100 | DEERWESTER S (1990), INDEXING BY LATENT SEMANTIC ANALYSIS, J AM SOC INFORM SCI, V41, P391 | 173 |
| 5 | BERNERSLEE T (2001), THE SEMANTIC WEB - A NEW FORM OF WEB CONTENT THAT IS MEANINGFUL TO COMPUTERS WILL UNLEASH A REVOLUTION OF NEW POSSIBILITIES, SCI AM, V284, P34 | 264 | MILNER R (1992), A CALCULUS OF MOBILE PROCESSES .1., INFORM COMPUT, V100, P1 | 95 | RAHM E (2001), A SURVEY OF APPROACHES TO AUTOMATIC SCHEMA MATCHING, VLDB J, V10, P334 | 100 |
| 6 | REITER R (1980), A LOGIC FOR DEFAULT REASONING, ARTIF INTELL, V13, P81 | 255 | DEERWESTER S (1990), INDEXING BY LATENT SEMANTIC ANALYSIS, J AM SOC INFORM SCI, V41, P391 | 94 | SMEULDERS AWM (2000), CONTENT-BASED IMAGE RETRIEVAL AT THE END OF THE EARLY YEARS, IEEE T PATTERN ANAL, V22, P1349 | 99 |
| 7 | MILNER R (1992), A CALCULUS OF MOBILE PROCESSES .1., INFORM COMPUT, V100, P1 | 245 | GELFOND M (1991), CLASSICAL NEGATION IN LOGIC PROGRAMS AND DISJUNCTIVE DATABASES, NEW GENERAT COMPUT, V9, P365 | 89 | MILLER GA (1995), WORDNET - A LEXICAL DATABASE FOR ENGLISH, COMMUN ACM, V38, P39 | 98 |
| 8 | GELFOND M (1991), CLASSICAL NEGATION IN LOGIC PROGRAMS AND DISJUNCTIVE DATABASES, NEW GENERAT COMPUT, V9, P365 | 240 | HAREL D (1987), STATECHARTS - A VISUAL FORMALISM FOR COMPLEX-SYSTEMS, SCI COMPUT PROGRAM, V8, P231 | 80 | GELFOND M (1991), CLASSICAL NEGATION IN LOGIC PROGRAMS AND DISJUNCTIVE DATABASES, NEW GENERAT COMPUT, V9, P365 | 97 |
| 9 | CHEN PPS (1976), THE ENTITY-RELATIONAL MODEL – TOWARD A UNIFIED VIEW OF DATA. ACM T DATABASE SYST, V1, P9 | 239 | MILLER GA (1995), WORDNET - A LEXICAL DATABASE FOR ENGLISH, COMMUN ACM, V38, P39 | 80 | GRUBER TR (1995), TOWARD PRINCIPLES FOR THE DESIGN OF ONTOLOGIES USED FOR KNOWLEDGE SHARING, INT J HUM-COMPUT ST, V43, P907 | 94 |
| 10 | VANEMDEN MH (1976), SEMANTICS OF PREDICATE LOGIC AS A PROGRAMMING LANGUAGE, J ASSOC COMPUT MACH, V23, P733 | 228 | KIFER M (1995), LOGICAL-FOUNDATIONS OF OBJECT-ORIENTED AND FRAME-BASED LANGUAGES, J ASSOC COMPUT MACH, V42, P741 | 71 | MILNER R (1992), A CALCULUS OF MOBILE PROCESSES .1., INFORM COMPUT, V100, P1 | 86 |
| 11 | HAREL D (1987), STATECHARTS - A VISUAL FORMALISM FOR COMPLEX-SYSTEMS, SCI COMPUT PROGRAM, V8, P231 | 210 | REITER R (1980), A LOGIC FOR DEFAULT REASONING, ARTIF INTELL, V13, P81 | 70 | GELFOND M (1988), THE STABLE MODEL SEMANTICS FOR LOGIC PROGRAMMING, P 5 INT C LOG PROGR, P1070 | 85 |
| 12 | CLARK KL (1978), NEGATION AS FAILURE, LOGIC DATA BASES, P293 | 191 | ASHBURNER M (2000), GENE ONTOLOGY: TOOL FOR THE UNIFICATION OF BIOLOGY, NAT GENET, V25, P25 | 67 | HORROCKS I (2003), FROM SHIQ AND RDF TO OWL: THE MAKING OF A WEB ONTOLOGY LANGUAGE, J WEB SEMANT, V1, P7 | 82 |
| 13 | ASHBURNER M (2000), GENE ONTOLOGY: TOOL FOR THE UNIFICATION OF BIOLOGY, NAT GENET, V25, P25 | 191 | GRUBER TR (1995), TOWARD PRINCIPLES FOR THE DESIGN OF ONTOLOGIES USED FOR KNOWLEDGE SHARING, INT J | 66 | VANGELDER A (1991), THE WELL-FOUNDED SEMANTICS FOR GENERAL LOGIC PROGRAMS, J ASSOC | 80 |

**Ying Ding**

| | | | HUM-COMPUT ST, V43, P907 | | COMPUT MACH, V38, P620 | |
|---|---|---|---|---|---|---|
| 14 | MILLER GA (1995), WORDNET - A LEXICAL DATABASE FOR ENGLISH, COMMUN ACM, V38, P39 | 187 | GIRARD JY (1987), LINEAR LOGIC, THEOR COMPUT SCI, V50, P1 | 61 | PORTER, M.F. (1980), AN ALGORITHM FOR SUFFIX STRIPPING, PROGRAM, V14, PP. 130-137 | 70 |
| 15 | GRUBER TR (1995), TOWARD PRINCIPLES FOR THE DESIGN OF ONTOLOGIES USED FOR KNOWLEDGE SHARING, INT J HUM-COMPUT ST, V43, P907 | 171 | USCHOLD M (1996), ONTOLOGIES: PRINCIPLES, METHODS AND APPLICATIONS, KNOWL ENG REV, V11, P93 | 55 | KALFOGLOU Y (2003), ONTOLOGY MAPPING: THE STATE OF THE ART, KNOWL ENG REV, V18, P1 | 65 |
| 16 | ZADEH LA (1965), FUZZY SETS, INFORM CONTR, V8, P338 | 155 | WIEDERHOLD G (1992), MEDIATORS IN THE ARCHITECTURE OF FUTURE INFORMATION-SYSTEMS, IEEE COMPUT, V25, P38 | 53 | LANDAUER TK (1998), AN INTRODUCTION TO LATENT SEMANTIC ANALYSIS, DISCOURSE PROCESS, V25, P259 | 63 |
| 17 | GIRARD JY (1987), LINEAR LOGIC, THEOR COMPUT SCI, V50, P1 | 148 | MCILRAITH SA (2001), SEMANTIC WEB SERVICES, IEEE INTELL SYST APP, V16, P46 | 51 | SEBASTIANI F (2002), MACHINE LEARNING IN AUTOMATED TEXT CATEGORIZATION, ACM COMPUT SURV, V34, P1 | 61 |
| 18 | HOARE CAR (1969), AN AXIOMATIC BASIS FOR COMPUTER PROGRAMMING, COMMUN ACM, V12, P576 | 142 | HENDLER J (2001), AGENTS AND THE SEMANTIC WEB, IEEE INTELL SYST APP, V16, P30 | 50 | MCILRAITH SA (2001), SEMANTIC WEB SERVICES, IEEE INTELL SYST APP, V16, P46 | 61 |
| 19 | MCCARTHY J (1980), CIRCUMSCRIPTION - A FORM OF NON-MONOTONIC REASONING, ARTIF INTELL, V13, P27 | 140 | SMEULDERS AWM (2000), CONTENT-BASED IMAGE RETRIEVAL AT THE END OF THE EARLY YEARS, IEEE T PATTERN ANAL, V22, P1349 | 49 | MILLER GA (1990), INTRODUCTION TO WORDNET: AN ON-LINE LEXICAL DATABASE, INT J LEXICOGR, V3, P235 | 60 |
| 20 | SMEULDERS AWM (2000), CONTENT-BASED IMAGE RETRIEVAL AT THE END OF THE EARLY YEARS, IEEE T PATTERN ANAL, V22, P1349 | 140 | MILLER GA (1990), INTRODUCTION TO WORDNET: AN ON-LINE LEXICAL DATABASE, INT J LEXICOGR, V3, P235 | 49 | DEMPSTER AP (1977), MAXIMUM LIKELIHOOD FROM INCOMPLETE DATA VIA EM ALGORITHM, J ROY STAT SOC B, V39, P1 | 58 |

Table 9 shows the highly cited papers from Scopus. As per Table 8, T. Gruber's ontology paper published in *Knowledge Acquisition* in 1993 again is ranked as the top one highly cited paper during all three periods. T. Berners-Lee, J. Hendler and O. Lassila's *Scientific American* journal article is ranked as the top two highly cited papers, Raymond Reiter's logic for default reasoning is ranked as the top two highly cited article in 2000-2004. S. Deerwester's latent semantic analysis from *Journal of the American Society for Information Science* is ranked as the top three highly cited paper in 1960-2009 and 2005-2009. There is no major difference between Table 8 and Table 9, even though WOS and Scopus have a significant different number of Semantic Web articles.

*Table 9. Highly cited papers (Scopus)*

| | 1960-2009 | | 2000-2004 | | 2005-2009 | |
|---|---|---|---|---|---|---|
| R | paper | no. cited | paper | no. cited | paper | no. cited |
| 1 | GRUBER TR (1993), A TRANSLATION APPROACH TO PORTABLE ONTOLOGY SPECIFICATIONS, KNOWL ACQUIS, V5, P199 | 598 | GRUBER TR (1993), A TRANSLATION APPROACH TO PORTABLE ONTOLOGY SPECIFICATIONS, KNOWL ACQUIS, V5, P199 | 121 | GRUBER TR (1993), A TRANSLATION APPROACH TO PORTABLE ONTOLOGY SPECIFICATIONS, KNOWL ACQUIS, V5, P199 | 470 |
| 2 | BERNERSLEE T (2001), THE SEMANTIC WEB - A NEW FORM OF WEB CONTENT THAT IS MEANINGFUL TO COMPUTERS WILL UNLEASH A REVOLUTION OF NEW POSSIBILITIES, SCI AM, V284, P34 | 416 | REITER, R. (1980), A LOGIC FOR DEFAULT REASONING, ARTIFICIAL INTELLIGENCE, V13, PP. 81-132 | 54 | BERNERSLEE T (2001), THE SEMANTIC WEB - A NEW FORM OF WEB CONTENT THAT IS MEANINGFUL TO COMPUTERS WILL UNLEASH A REVOLUTION OF NEW POSSIBILITIES, SCI AM, V284, P34 | 355 |
| 3 | DEERWESTER, S. (1990), INDEXING BY LATENT SEMANTIC ANALYSIS, JOURNAL OF THE AMERICAN SOCIETY FOR INFORMATION | 132 | VAN GELDER, A. (1991), THE WELL-FOUNDED SEMANTICS FOR GENERAL LOGIC PROGRAMS, JOURNAL OF THE | 52 | DEERWESTER, S. (1990), INDEXING BY LATENT SEMANTIC ANALYSIS, JOURNAL OF THE AMERICAN | 93 |

# Ying Ding

| | | | | | | | |
|---|---|---|---|---|---|---|---|
| | SCIENCE, V41, PP. 391-407 | | ACM, V38, PP. 620-650 | | SOCIETY FOR INFORMATION SCIENCE, V41, PP. 391-407 | |
| 4 | GELFOND M (1991), CLASSICAL NEGATION IN LOGIC PROGRAMS AND DISJUNCTIVE DATABASES, NEW GENERAT COMPUT, V9, P365 | 110 | HAREL D (1987), STATECHARTS - A VISUAL FORMALISM FOR COMPLEX-SYSTEMS, SCI COMPUT PROGRAM, V8, P231 | 47 | LANDAUER, T.K. (1998), AN INTRODUCTION TO LATENT SEMANTIC ANALYSIS, DISCOURSE PROCESSES, V25, PP. 259-284 | 71 |
| 5 | REITER, R. (1980), A LOGIC FOR DEFAULT REASONING, ARTIFICIAL INTELLIGENCE, V13, PP. 81-132 | 104 | BERNERSLEE T (2001), THE SEMANTIC WEB - A NEW FORM OF WEB CONTENT THAT IS MEANINGFUL TO COMPUTERS WILL UNLEASH A REVOLUTION OF NEW POSSIBILITIES, SCI AM, V284, P34 | 46 | KALFOGLOU, Y. (2003), ONTOLOGY MAPPING: THE STATE OF THE ART, THE KNOWLEDGE ENGINEERING REVIEW, V18, PP. 1-31 | 71 |
| 6 | LANDAUER, T.K. (1998), AN INTRODUCTION TO LATENT SEMANTIC ANALYSIS, DISCOURSE PROCESSES, V25, PP. 259-284 | 89 | GELFOND M (1991), CLASSICAL NEGATION IN LOGIC PROGRAMS AND DISJUNCTIVE DATABASES, NEW GENERAT COMPUT, V9, P365 | 41 | SEBASTIANI F (2002), MACHINE LEARNING IN AUTOMATED TEXT CATEGORIZATION, ACM COMPUT SURV, V34, P1 | 68 |
| 7 | HAREL D (1987), STATECHARTS - A VISUAL FORMALISM FOR COMPLEX-SYSTEMS, SCI COMPUT PROGRAM, V8, P231 | 88 | DEERWESTER, S. (1990), INDEXING BY LATENT SEMANTIC ANALYSIS, JOURNAL OF THE AMERICAN SOCIETY FOR INFORMATION SCIENCE, V41, PP. 391-407 | 39 | GELFOND M (1991), CLASSICAL NEGATION IN LOGIC PROGRAMS AND DISJUNCTIVE DATABASES, NEW GENERAT COMPUT, V9, P365 | 68 |
| 8 | RAHM, E. (2001), A SURVEY OF APPROACHES TO AUTOMATIC SCHEMA MATCHING, VLDB JOURNAL, V10, PP. 334-350 | 87 | KRAUS, S. (1990), NONMONOTONIC REASONING, PREFERENTIAL MODELS AND CUMULATIVE LOGICS, ARTIFICIAL INTELLIGENCE, V44, PP. 167-207 | 35 | HORROCKS I (2003), FROM SHIQ AND RDF TO OWL: THE MAKING OF A WEB ONTOLOGY LANGUAGE, J WEB SEMANT, V1, P7-26 | 66 |
| 9 | SEBASTIANI F (2002), MACHINE LEARNING IN AUTOMATED TEXT CATEGORIZATION, ACM COMPUT SURV, V34, P1 | 86 | ZADEH, L.A. (1965), FUZZY SETS, INFORMATION AND CONTROL, V8, PP. 338-353 | 34 | TVERSKY, A. (1977), FEATURES OF SIMILARITY, PSYCHOLOGICAL REVIEW, V84, PP. 327-352 | 65 |
| 10 | PORTER, M.F. (1980), AN ALGORITHM FOR SUFFIX STRIPPING, PROGRAM, V14, PP. 130-137 | 84 | GIRARD JY (1987), LINEAR LOGIC, THEOR COMPUT SCI, V50, P1 | 32 | RAHM, E. (2001), A SURVEY OF APPROACHES TO AUTOMATIC SCHEMA MATCHING, VLDB JOURNAL, V10, PP. 334-350 | 62 |
| 11 | KALFOGLOU, Y. (2003), ONTOLOGY MAPPING: THE STATE OF THE ART, THE KNOWLEDGE ENGINEERING REVIEW, V18, PP. 1-31 | 79 | MESEGUER, J. (1992), CONDITIONAL REWRITING LOGIC AS A UNIFIED MODEL OF CONCURRENCY, THEORETICAL COMPUTER SCIENCE, V96, PP. 73-155 | 29 | STUDER, R. (1998), KNOWLEDGE ENGINEERING: PRINCIPLES AND METHODS, DATA AND KNOWLEDGE ENGINEERING, V25, PP. 161-197 | 59 |
| 12 | TVERSKY, A. (1977), FEATURES OF SIMILARITY, PSYCHOLOGICAL REVIEW, V84, PP. 327-352 | 79 | MCCARTHY J (1980), CIRCUMSCRIPTION - A FORM OF NON-MONOTONIC REASONING, ARTIF INTELL, V13, P27 | 29 | ZHUGE, H. (2004), CHINA'S E-SCIENCE KNOWLEDGE GRID ENVIRONMENT, IEEE INTELLIGENT SYSTEMS, V19, PP. 13-17 | 58 |
| 13 | ALUR, R. (1994), A THEORY OF TIMED AUTOMATA, THEORETICAL COMPUTER SCIENCE, V126, PP. 183-235 | 79 | PORTER, M.F. (1980), AN ALGORITHM FOR SUFFIX STRIPPING, PROGRAM, V14, PP. 130-137 | 28 | ALUR, R. (1994), A THEORY OF TIMED AUTOMATA, THEORETICAL COMPUTER SCIENCE, V126, PP. 183-235 | 58 |
| 14 | ZADEH, L.A. (1965), FUZZY SETS, INFORMATION AND CONTROL, V8, PP. 338-353 | 78 | JACOBS, B. (1997), A TUTORIAL ON (CO)ALGEBRAS AND (CO)INDUCTION | 25 | USCHOLD, M. (1996), ONTOLOGIES: PRINCIPLES, METHODS AND APPLICATIONS, KNOWLEDGE ENGINEERING REVIEW, V11, PP. 93-136 | 57 |
| 15 | USCHOLD, M. (1996), ONTOLOGIES: PRINCIPLES, METHODS AND APPLICATIONS, KNOWLEDGE ENGINEERING REVIEW, V11, PP. 93-136 | 75 | RAHM, E. (2001), A SURVEY OF APPROACHES TO AUTOMATIC SCHEMA MATCHING, VLDB JOURNAL, V10, PP. 334-350 | 25 | PORTER, M.F. (1980), AN ALGORITHM FOR SUFFIX STRIPPING, PROGRAM, V14, PP. 130-137 | 56 |

**Ying Ding**

| 16 | STUDER, R. (1998), KNOWLEDGE ENGINEERING: PRINCIPLES AND METHODS, DATA AND KNOWLEDGE ENGINEERING, V25, PP. 161-197 | 74 | COHEN, P.R. (1990), INTENTION IS CHOICE WITH COMMITMENT, ARTIFICIAL INTELLIGENCE, V42, PP. 213-261 | 23 | DEERWESTER, S.C. (1990), INDEXING BY LATENT SEMANTIC ANALYSIS, JOURNAL OF THE AMERICAN SOCIETY OF INFORMATION SCIENCE, V41, PP. 391-407 | 55 |
|---|---|---|---|---|---|---|
| 17 | VANGELDER A (1991), THE WELL-FOUNDED SEMANTICS FOR GENERAL LOGIC PROGRAMS, J ASSOC COMPUT MACH, V38, P620 | 73 | HENDLER J (2001), AGENTS AND THE SEMANTIC WEB, IEEE INTELL SYST APP, V16, P30-37 | 23 | RESNIK, P. (1999), SEMANTIC SIMILARITY IN A TAXONOMY: AN INFORMATION-BASED MEASURE AND ITS APPLICATION TO PROBLEMS OF AMBIGUITY IN NATURAL LANGUAGE, JOURNAL OF ARTIFICIAL INTELLIGENCE RESEARCH, V11, PP. 95-130 | 54 |
| 18 | HENDLER J (2001), AGENTS AND THE SEMANTIC WEB, IEEE INTELL SYST APP, V16, P30-37 | 72 | MILNER, R. (1992), A CALCULUS OF MOBILE PROCESSES, INFORMATION AND COMPUTATION, V100, PP. 1-77 | 44 | RAHM, E. (2001), A SURVEY OF APPROACHES TO AUTOMATIC SCHEMA MATCHING, VLDB JOURNAL, V10, PP. 334-350 | 54 |
| 19 | LANDAUER, T.K. (1997), A SOLUTION TO PLATO'S PROBLEM: THE LATENT SEMANTIC ANALYSIS THEORY OF ACQUISITION, INDUCTION, AND REPRESENTATION OF KNOWLEDGE, PSYCHOLOGICAL REVIEW, V104, PP. 211-240 | 71 | MILLER, G.A. (1995), WORDNET - A LEXICAL DATABASE FOR ENGLISH, COMMUNICATIONS OF THE ACM, V38, PP. 39-41 | 23 | LANDAUER, T.K. (1997), A SOLUTION TO PLATO'S PROBLEM: THE LATENT SEMANTIC ANALYSIS THEORY OF ACQUISITION, INDUCTION, AND REPRESENTATION OF KNOWLEDGE, PSYCHOLOGICAL REVIEW, V104, PP. 211-240 | 51 |
| 20 | HORROCKS I (2003), FROM SHIQ AND RDF TO OWL: THE MAKING OF A WEB ONTOLOGY LANGUAGE, J WEB SEMANT, V1, P7-26 | 70 | MOGGI, E. (1991), NOTIONS OF COMPUTATION AND MONADS, INFORMATION AND COMPUTATION, V93, PP. 55-92 | 23 | HENDLER J (2001), AGENTS AND THE SEMANTIC WEB, IEEE INTELL SYST APP, V16, P30-37 | 49 |

These top highly cited papers from WOS and Scopus can be grouped into different schools-of-thought:

- *Vision*: T. Berners-Lee's "The semantic web".
- *Ontology Engineering*: T. Gruber's "A translation approach to portable ontology specifications"; T. Gruber's "Toward principles for the design of ontologies used for knowledge sharing"; M. Uschold's "Ontologies: principles, methods and applications"; Y. Kalfouglou's "Ontology mapping: the state of the art" and R. Studer's "Knowledge engineering: Principles and methods".
- *Ontological Languages*: I. Horrocks' "From SHIQ and RDF to OWL".
- *Semantic Web Services*; S. Mcilraith's "Semantic web services"; J. Hendler's "Agents and the semantic web" and H. Zhuge's "China's E-science knowledge grid environment".
- *Core Artificial Intelligence*: M. Gelfond's "The stable model semantics for logic programming"; A. van Gelder's "The well-founded semantics for general logic programs"; R. Reiter's "A logic for default reasoning"; R. Milner's "A calculus of mobile processes"; M. Gelfond's "Classical negation in logic programs and disjunctive databases"; M. Van Emden's "Semantics of predicate logic as a programming language"; K. Clark's "Negation as failure"; L. Zadeh's "Fuzzy sets"; J. Girard's "Linear logic"; C. Hoare's "An axiomatic basis for computer programming"; J. McCarthy's "Circumscription – A form of non-monotonic reasoning" M. Kifer's "Logical foundations of object-oriented and frame-based languages"; R. Alur's "A theory of timed automata"; S. Kraus' "Non-monotonic reasoning, preferential models and cumulative logic"; J. Meseguer's "Conditional rewriting logic as a unified model of concurrency"; B. Jacobs' "A tutorial on (co)algebras and (co)induction"; P. Cohen's "Intention is choice with commitment"; E. Moggi's "Notions of computation and monads" and D. Harel's "Statecharts – A visual formalism for complex systems."

- *Related fields:*
    - *Information Retrieval*: S. Deerwester's "Indexing by latent semantic analysis": A. Smeulders' "Content-based image retrieval at the end of the early years"; T. Landauer's "An introduction to latent semantic analysis" and A. Tversky's "Features of similarity".
    - *Database*: P. Chen's "The entity-relational model"; G. Wiederhold's "Mediators in the architecture of future information-systems" and E. Rahme's A survey of approaches to automatic schema matching.
    - *Bioinformatics*: M. Ashburner's "Gene ontology".
    - *Natural Language Processing*: G. Miller's "Wordnet-A lexical database for English"; M. Porter's "An algorithm for suffix stripping" and P. Resnik's "Semantic similarity in a taxonomy: An information-based measure and its application to problems of ambiguity in natural language".
    - *Data/text Mining*: F. Sebastiani's "Machine learning in automated text categorization" and A. Dempster's "Maximum likelihood from incomplete data via EM algorithm".

These highly cited papers in the related fields do not belong to semantic web area, but they are highly cited articles by the semantic web researchers. For example, S. Deerwester's "Indexing by latent semantic analysis" is one of the best algorithms to derive topics therefore forms the fundamental methods for ontology learning. Similar for highly cited papers in database and mediator (as RDF triple stores are related to database), text mining and Natural Language Processing (as they are the major building blocks for ontology learning and mapping), and bioinformatics (as it is one of the leading areas which apply semantic web technologies and achieve appealing results).

### 5.3 New stars in the Semantic Web

Table 10 shows the top 20 authors with the highest increase of their citations from 2000-2004 to 2005-2009. In WOS, M. A. Harris (Gene Ontology-related research), T. Harris (design and implementation of programming languages) and L. Ding (Swoogle – Semantic Web Search Engine) are ranked as the top three authors with the highest increase of citations. Coming from Scopus, D. Roman (Semantic Web Services), J. De Bruijn (logic programming) and L. Ding (Swoogle) are ranked as top three for the significant increase in number of citations.

Table 10. New stars

|   | Web of Science |                  | Scopus     |                  |
|---|----------------|------------------|------------|------------------|
| R | Name           | Times of increase | Name       | Times of increase |
| 1 | HARRIS MA      | 30.5             | Roman, D.  | 72.5             |
| 2 | HARRIS T       | 21.5             | De Bruijn, J. | 70            |
| 3 | DING L         | 20.7             | Ding, L.   | 43               |
| 4 | MARCUS A       | 20.5             | Harris, T. | 37.5             |
| 5 | ROMAN D        | 19               | Rao, J.    | 36               |
| 6 | CHEN YX        | 18.5             | Carroll, J.J. | 35            |
| 7 | ANTONIOL G     | 17.5             | Hollink, L. | 34              |
| 8 | HAASE P        | 16.8             | Monay, F.  | 32               |
| 9 | KNUBLAUCH H    | 16               | Lara, R.   | 30.5             |
| 10 | ALSHAHROUR F  | 15.3             | Tang, J.   | 29               |
| 11 | JEON J        | 15               | Gu, T.     | 28.5             |
| 12 | LIERLER Y     | 14               | Haase, P.  | 27.6             |
| 13 | LARA R        | 14               | Bowers, S. | 27.5             |
| 14 | DONNELLY M    | 14               | Gauch, S.  | 27.5             |

| 15 | PATWARDHAN S | 14 | Snoek, C.G.M. | 27.25 |
| --- | --- | --- | --- | --- |
| 16 | PRUDHOMMEAUX E | 13.6 | Rosati, R. | 26.7 |
| 17 | MA YF | 13 | Pang, B. | 26.5 |
| 18 | PANTEL P | 12.7 | Prud'hommeaux, E. | 26 |
| 19 | WANG P | 12.3 | Ding, Z. | 25.5 |
| 20 | FU X, MAXIMILIEN EM, VENNEKENS J | 12 | Akkiraju, R. | 25 |

Notes: Times of increase = [(No. of being cited in 2005-2009)-(No. of being cited in 2000-2004)]/(No. of being cited in 2000-2004)

## 6. Conclusion

This paper conducted citation analysis for the field of Semantic Web covering 1960-2009. Papers and citations were collected from two major databases, Web of Science and Scopus. The productivity and impact of the Semantic Web community have been analyzed, notably within the last decade of development for the periods of 2000-2004 and 2005-2009. The major publication channels in the Semantic Web field are conference proceedings, especially those published by Springer as the series *Lecture Notes in Computer Sciences*. Major journals that publish Semantic Web papers are *Theoretical Computer Science*, *Bioinformatics*, *Data and Knowledge Engineering* and *IEEE Transactions on Knowledge and Data Engineering*. The most productive authors are T. Eiter, A. Brogi and H. Zhuge. J. J Jung is the newly emerging, very productive author in this field.

The research impact has been analyzed based on citation counting. In the whole period (1960-2009), R. Milner, M. Gelfond and C. A. R. Hoare are ranked as the top three authors. Sir Tim Berners-Lee is ranked fourth throughout the period. Scopus citation data allows the ranking of cited second or third authors. J. Hendler, S. Staab and H. Garcia-Molina are ranked as the top three highly cited second authors, while O. Lassila, F. van Harmelen and I. Horrocks are the top three highly cited third authors. In WOS, T. Gruber's ontology paper has been consistently highly cited and ranked top for all sub-periods. A. Van Gelder's theory proving paper is ranked two, and S. Deerwester's latent semantic analysis paper is ranked three. T. Berners-Lee, J. Hendler and O. Lassila's article about the vision of the Semantic Web, published in *Scientific American*, is ranked as the top second highly cited paper in 2005-2009, while in Scopus, Gruber's ontology paper and Berners-Lee's Scientific American papers are the top two highly cited papers in 1960-2009 and 2005-2009. In both WOS and Scopus, the highly cited journals and conferences are *Lecture Notes in Computer Science* or *Lecture Notes in Artificial Intelligence*, *Artificial Intelligence*, *Communication of the ACM* and *Theoretical Computer Science*. In WOS, M. A. Harris, T. Harris and L. Ding are ranked as the top three authors with the highest increase of citations, while from Scopus, D. Roman, J. De Bruijn and L. Ding are ranked as the top three for the significant increase in number of citations.

By comparing highly cited articles in 2000-2004 and 2005-2009, one can see the research shifting from core AI-related logic programming, logic reasoning and theory proving, to ontological languages (e.g., RDF, OWL), semantic data conversion and ontology mapping. One may therefore predict that within the next ten years, the following topics may become mainstreams in this field:

- *Creating, converting and enriching semantic data*: this mainstream effort is led by the Linked Open Data (LOD) Initiatives created by C. Bizer of Free University of Berlin, Germany. LOD bubbles will grow to an amazing degree, becoming the major showcase of Semantic Web technologies. LOD creates the test bed for semantic query, reasoning and service/data mashups. It demonstrates a powerful, simple, flexible and efficient approach to integrating heterogeneous datasets and triggers the industrial, governmental and academic adoption. In 2010-2020, the efforts might focus on the quality issue of the LOD data, scalability of managing and querying LOD data, and security on data and SPARQL query;
- *Mining semantic RDF/OWL graphs*: Semantic Web creates better technologies to represent and integrate data, while all these efforts should lead to the final goal: providing better search technologies. Since RDF

**Ying Ding**

    data form graphs, the searching and retrieving of RDF data utilizes the current Google approach: PageRank or HITS, wherein the topologies of graphs play the major role in ranking nodes in the networks. RDF graphs contain more semantics than normal graphs in Google, as the links and nodes are instances of the ontologies. Various weighted, topic-sensitive or semantic-sensitive PageRank may therefore become a new research topic in the ranking of semantic nodes. Provenance data once again becomes meaningful, wherein datasets need to be integrated. This development traces different steps of data integration and enables provenance-based layered data analysis, query and visualization;

- *Simple reasoning*: Revolutionary breakthroughs should happen during the next few years as complex reasoning fails to scale up. Reasoning should be kept as simple as possible, scalable and error-tolerant. Relaxed or simplified logic may thus be invented to make this fly.
- *Benchmarking and evaluating ontologies*: Nowadays ontologies have been created nearly everywhere, as noted in the introduction – a necessary step for solving the information-deluge problem. There is a pressing need to create a benchmark or widely adopted framework to evaluate and test these ontologies. Notably during the process of generating ontologies, domain experts may have a handbook in hand to ensure right decisions on the modeling of their classes, properties or instances. Examples may be found from other communities, such as TREC in information retrieval;
- *Interfacing Semantic Web*: The next ten years should see to the creation of an innovative user-friendly interface to showcase the Semantic Web. Actually achieving goals of the Semantic Web is still currently impossible, as the search interface or SPARQL Endpoints for LOD datasets are not really targeted for normal users, and are instead accessible to SW gurus or hackers. To bring the Semantic Web out of the research lab and make its debut for normal users, a simple interface design is essential;
- *Utilizing social Web (Web2.0)*: The current social network fever in Web2.0 facilitates the generation of social semantic data, such as social tagging, commenting voting and recommending. These data identify existing relationships and create new ones, forming a "social power" that helps the LOD community snowball their datasets and introduce mashup powers of Semantic Web technologies. In the next ten years, we may predict that Web2.0 and the Semantic Web will be merged or interwoven in the manner that motivates normal Web2.0 users to contribute more social metadata, while Semantic Web should provide better technologies to mashup these data and further stimulate data generation. The difference between Web2.0 and Semantic Web will become blurred, as they finally merge to become the next generation Web – Web3.0 – which extends current Web2.0 applications using Semantic Web technologies and graph-based open data [26];
- *Embracing eScience and eGovernment*: In the next ten years, eScience and eGovernment will be the major adopters of Semantic Web technologies. The current trend toward data integration, interlinking and analysis within health sciences, biology, medicine, pharmaceuticals and chemistry will lead to new technologies such as bio2rdf, Linked Open Drug Data and YeastHub. Semantic publishing will create new norms for the next generation of publishing, where RDF triples will be asked to add to paper during the submission required by the publishers like the current authors are all familiar with adding keywords to their articles. Journal or conference papers are no long just pure "static strings". They contain important RDF triples which are interlinked in the paper, with other related papers (e.g., citations), and outside related semantic datasets (e.g., LOD bubbles). The substantial funding secured from NIH for CTSA[8] and research networking for life science indicates the confidence and uptake of the Semantic Web technologies from other major funding agents in the United States, including NSF. The recent groundbreaking news from the USA and UK that their governments are ready to use the potential of the

---

[8] http://www.ncrr.nih.gov/clinical_research_resources/clinical_and_translational_science_awards/

**Ying Ding**

Semantic Web technologies to build their transparent eGovernement platform (Shadbolt, Hall & Berners-Lee, 2006). These, by no doubt, will create the tremendous momentum and broad social and societal impact on the Semantic Web. This momentum will radiate other fields which data integration is essential, such as environmental science to integrate data from hydrology, climatology, ecology, and oceanography [27].

The challenges to the Semantic Web may be as significant as its promises. As I. Horrocks mentioned in his recent article, "*The vision of a Semantic Web is extremely ambitious and would require solving many long-standing research problems in knowledge representation and reasoning, databases, computational linguistics, computer vision, and agent system.*" [8]. To carry on and further realize this vision, the Semantic Web community needs to work with researchers from related fields to establish the Semantic Web as the emerging interdisciplinary field – called "Web Science" – to view the World Wide Web as an important entity to be studied in its own right, and to understand its future as a computational structure and an interacting platform of people and machine [9].

Although there are twice as many Semantic Web papers in Scopus as those in WOS, the citation analysis for the field of Semantic Web does not show a significant difference between the two. For future research, we plan to use social network analysis to detect research groups or communities in this field. The use of self-citation also poses a new area of research that can be further extended to group self-citation or project self-citation in papers citing or cited by authors from the same research group or related projects. This may help identify the knowledge diffusion and transfer patterns in this field, as new and existing thinkers within this closely-knit community become necessarily self-referential.

# Ying Ding